\documentclass[prd,aps,preprintnumbers,nofootinbib,superscriptaddress,notitlepage,floatfix,11pt]{revtex4}
\usepackage{stmaryrd}
\usepackage{epsfig}
\usepackage{amsfonts}
\usepackage{amsmath}
\usepackage{graphicx}
\usepackage{color}
\usepackage{amssymb,amsmath,psfrag,slashed,graphicx}
\usepackage[normalem]{ulem}

\def \nn {\nonumber}

\def\slash#1{#1 \hskip-0.45em /}

\allowdisplaybreaks

\begin{document}
\title{Matching the $B$-meson quasidistribution amplitude in the RI/MOM scheme}

\author{Ji Xu~\footnote{xuji\_phy@zzu.edu.cn}}
\affiliation{School of Physics and Microelectronics, Zhengzhou University, Zhengzhou, Henan 450001, China}

\author{Xi-Ruo Zhang~\footnote{ZXRxiruo@163.com}}
\affiliation{School of Physics and Microelectronics, Zhengzhou University, Zhengzhou, Henan 450001, China}

\begin{abstract}
  Within the framework of large momentum effective theory (LaMET), the light-cone distribution amplitude of $B$-meson in heavy-quark effective theory (HQET) can be extracted from lattice calculations of quasidistribution amplitude through hard-collinear factorization formula. This quasiquantity can be renormalized in a regularization-independent momentum subtraction scheme (RI/MOM). In this work, we derive the matching coefficient which connects the renormalized quasiditribution amplitude in the RI/MOM scheme and standard LCDA in the $\overline{\textrm{MS}}$ scheme at one-loop accuracy. Our numerical analysis approves of the feasibility of RI/MOM scheme for renormalizing $B$-meson quasidistribution amplitude. These results will be crucial for exploring the partonic structure of heavy-quark hadrons.
\end{abstract}

\maketitle

\section{Introduction}
\label{sec:Introduction}
$B$-meson light-cone distribution amplitudes (LCDAs) in heavy-quark effective theory (HQET) are the most basic objects about the structure of this hadron with which the QCD factorization theorems of exclusive $B$-meson decay become experimentally verifiable \cite{Grozin:1996pq,Beneke:1999br,Beneke:2000wa,Beneke:2001at,Becher:2005fg}. Defined as the light-cone matrix elements of the nonlocal HQET quark-gluon operators, they describe the nonperturbative strong interaction dynamics of the $B$-meson system. Although there have been many progresses in perturbative calculations concerning $B$-meson decays in recent years \cite{Wang:2016beq,Wang:2016qii,Feldmann:2014ika,Bell:2013tfa,Galda:2022dhp,Deng:2021zoi,Zhao:2019elu,Yao:2022zom}, our limited knowledge of $B$-meson LCDAs has become the major stumbling block for precision predictions of the $B$-meson decay observables. Therefore currently, the point significant in $B$ physics is improving the accuracy of $B$-meson LCDAs.

Despite its importance, calculating LCDAs from first principles of quantum chromodynamics (QCD) has been a challenge. Model-independent properties of the leading-twist $B$-meson LCDA $\phi_B^+(\omega, \mu)$ and its first inverse moment $\lambda_B^{-1}(\mu)$ have received considerable amount of attention lately \cite{Lange:2003ff,Braun:2019wyx,Lee:2005gza,Wang:2018wfj,Beneke:2018wjp}. By contrast, nonperturbative determinations of $\phi_B^+(\omega, \mu)$ have been mainly performed in the framework of QCD sum rules (QCDSR) or Dyson-Schwinger equation (DSE) \cite{Braun:2003wx,Gao:2014bca}, whereas both theories have their own drawbacks. For the former, it lies in the fact that the light-cone separation between the effective heavy-quark field and the light antiquark field needs to be sufficiently small to guarantee the validity of the local operator product expansion (OPE) for the HQET correlation function under discussion. For the latter, the DSEs are essentially equations of motion corresponding to the Green's function whose solution requires accurate knowledge of the $B$-meson wave function. Therefore, it is then evident that determining the momentum dependence of $B$-meson LCDAs with model-independent techniques is of top priority in $B$ physics. Being nonperturbative in nature, LCDAs intrinsically contain low energy degrees of freedom thus cannot be evaluated in perturbation theory. Nonperturbative methods such as lattice QCD offers an alternative way out. However, the dependence of LCDA correlator on the light-cone coordinate makes it essentially unfeasible to be directly calculated on the lattice which is constructed in Euclidean space with imaginary time.

A promising approach to circumvent this problem has been proposed under the name of large momentum effective theory (LaMET) \cite{Ji:2013dva,Ji:2014gla}. The essential strategy of this novel proposal resides in the construction of a time-independent quasiquantity which, on the one hand, can be readily computed on a Euclidean lattice and, on the other hand, approaches the original hadronic distribution amplitude on the light-cone under Lorentz boost. The fairly encouraging results from the state-of-art computations of the nucleon PDFs and the light-meson distribution amplitudes evidently demonstrate that the LaMET formalism allows for a bright future to systematically compute a wide range of ``parton observables'' with the demanding computational resources and the tremendous development of new techniques and algorithms \cite{Ji:2015qla,Xiong:2015nua,Li:2016amo,Ishikawa:2016znu,Monahan:2016bvm,Constantinou:2017sej,Ji:2017oey,Jia:2017uul,Wang:2017qyg,Stewart:2017tvs,Wang:2017eel,Xu:2018mpf,Xu:2018eii,Radyushkin:2018nbf,Zhang:2018diq,Li:2018tpe,Liu:2019urm,Liu:2018tox,Ji:2020brr,
Hua:2022kcm,Hua:2022wop,Hua:2020gnw,Ji:2020ect,Cichy:2018mum,Bhat:2022zrw,Egerer:2021ymv,Alexandrou:2020uyt,Alexandrou:2020zbe,Bhattacharya:2020xlt,Su:2022fiu}.
In view of the significance of $B$-meson LCDAs and the validity of LaMET, proposing approaches to determine $B$-meson LCDAs in the frame of LaMET is a matter to which people should attach much more attentions and there have been some preliminary researches \cite{Wang:2019msf,Kawamura:2018gqz,Xu:2022krn}.

Based on LaMET, the procedure of calculating $B$-meson LCDA from lattice QCD can be divided into three steps. 1. Lattice simulation on the $B$-meson quasidistribution amplitude; 2. Renormalizing the quasidistribution amplitude in a specific scheme; 3. Matching the renormalized quasidistribution amplitude to LCDA which is usually renormalized in the $\overline{\textrm{MS}}$ scheme. In this paper, we focus on the second and third steps. With increasing computational resources, the renormalization process will be a key factor to improve the precision of $B$-meson quasidistribution amplitude. The authors in \cite{Wang:2019msf} constructed the quasidistribution amplitude $\varphi_B^+(\xi,\mu)$ and renormalized it in the $\overline{\textrm{MS}}$ scheme. One of the standard methods to renormalize operators in lattice QCD is regularization-independent momentum subtraction (RI/MOM) scheme which essentially belongs to momentum subtraction schemes in quantum field theory. As a nonperturbative method, it has proven to be practical in the frame of LaMET and gained great popularity in recent years \cite{Alexandrou:2017huk,Chen:2017mzz,Lin:2017ani,Green:2017xeu} (see \cite{Radyushkin:2017cyf,Orginos:2017kos,Braun:2018brg,Li:2020xml} for other practical approaches). The multiplicative renormalizability of the constructed quasi-HQET operator to all orders in perturbation theory has been demonstrated, which enables a nonperturbative renormalization such as RI/MOM scheme. This is a crucial step in the application of extracting $B$-meson LCDA in lattice. After being renormalized in RI/MOM scheme, then the $B$-meson quasidistribution amplitude can be matched onto the usual $B$-meson LCDA through factorization formula. A perturbative matching coefficient appearing in the formula that converts the $B$-meson quasidistribution amplitude in the RI/MOM scheme to $B$-meson LCDA in the $\overline{\textrm{MS}}$ scheme is not available yet. One of the main motives in this paper is to calculating this coefficient at one-loop accuracy.

Our work is an extension of a series of previous works. The $B$-meson quasidistribution amplitude $\varphi_B^+(\xi,\mu)$ renormalized in the RI/MOM scheme and the perturbative matching coefficient entering the hard-collinear factorization formula will be presented. Since the renormalized matrix elements in the RI/MOM scheme are independent on UV regularization choices, we carry out this matching calculation with dimensional regularization for convenience. These results will be crucial to exploring the partonic structure of heavy-quark hadrons in the static limit.

The rest of this paper is organized as follows: In Sec.\,\ref{sec:Review}, the leading twist (twist-2) LCDA and quasi-DA as well as the RI/MOM scheme will be briefly reviewed. Then in Sec.\,\ref{sec:Calculation}, we present the factorization formula, followed by the calculation of the renormalized quasidistribution amplitude and the derived matching coefficient. In Sec.\,\ref{sec:Numerical} we analyze these results and give perspectives for lattice calculations, a numerical comparison between the $B$-meson quasidistribution amplitude obtained in the RI/MOM scheme and a modeled $B$-meson LCDA would be presented. We conclude in Sec.\,\ref{sec:Conclusion}. A few more details about calculation of renormalized quasidistribution amplitude are left to Appendix.

\section{$B$-meson (quasi)Distribution amplitudes and RI/MOM scheme}
\label{sec:Review}
The momentum space distribution function of leading-twist LCDA $\phi_B^+(\omega,\mu)$ can be deduced from Fourier transformation of its form in coordinate space \cite{Braun:2003wx}
\begin{eqnarray}\label{LCDA}
  \phi_{B}^{+}(\omega, \mu)=\frac{1}{2 \pi} \int_{-\infty}^{+\infty} d \eta \, e^{i \bar{n} \cdot v  \omega  \eta} \tilde{\phi}_{B}^{+}(\eta-i \epsilon, \mu) \,,
\end{eqnarray}
here $\bar n$ is the light-cone coordinate with $\bar n^2=0$, and $\tilde \phi_B^+(\eta,\mu)$ is the leading-twist LCDA in coordinate space with the definition
\begin{eqnarray}\label{LCDA2}
 \langle 0 |(\bar{q} W_{c})(\eta \bar{n}) \slash{\bar n} \gamma_{5} (W_{c}^{\dagger} h_{v})(0) | \bar{B}(v) \rangle = i \tilde{f}_{B}(\mu) m_{B} \tilde{\phi}_{B}^{+}(\eta, \mu) \,.
\end{eqnarray}
The soft light-cone Wilson line is given by $W_{c}(\eta \bar{n})=\mathrm{P}\left\{\operatorname{Exp}\left[i g_{s} \int_{-\infty}^{\eta} d x \, \bar{n} \!\cdot\! A(x \bar{n})\right]\right\}$ and $\tilde{f}_{B}(\mu)$ is the static decay constant of $B$-meson \cite{Beneke:2005gs}.

We will employ the following definition of $B$-meson quasidistribution amplitude
\begin{eqnarray}\label{quasiDA}
  i \tilde{f}_{B}(\tilde\mu) m_{B} \varphi_{B}^{+}(\xi, \tilde\mu) = \int_{-\infty}^{+\infty} \frac{d \tau}{2 \pi} e^{i n_{z} \cdot v \xi \tau}\langle 0| (\bar{q} W_{c})(\tau n_{z}) \slash{n}_{z} \gamma_{5}(W_{c}^{\dagger} h_{v})(0) | \bar{B}(v) \rangle \,.
\end{eqnarray}
Here $\tilde\mu$ is a renormalization scale for the quasidistribution amplitude whose definition depends on the renormalization scheme we choose. One can see that $\varphi_{B}^{+}(\xi, \tilde\mu)$ is constructed by the spatial correlation function of two collinear (effective) quark fields with $n_z=(0,0,0,1)$ and we will work in a Lorentz boosted frame of the $B$-meson in which $\bar n \cdot v \gg n\cdot v$ and set $v_{\perp\mu}=0$. Unlike the $B$-meson LCDA defined in Eq.\,(\ref{LCDA}), which is invariant under a boost along the $z$ direction, the quasidistribution amplitude changes dynamically under such a boost, which is encoded in its nontrivial dependence on the heavy quark velocity $v$.

It is of vital importance to show that the nonlocal matrix element in Eq.\,(\ref{quasiDA}) will renormalize multiplicatively to all orders in perturbation theory applying the lattice regularization scheme since this feature will facilitate the lattice QCD simulation substantially. \cite{Wang:2019msf} has demonstrated this multiplicative renormalizability which enables the RI/MOM scheme to be utilized in the $B$-meson quasidistribution amplitude $\varphi_{B}^{+}(\xi, \tilde\mu)$. Following the strategy in \cite{Liu:2019urm,Liu:2018tox}, the RI/MOM renormalization factor $Z_{\textrm{OM}}$ is determined nonperturbatively on lattice by imposing the condition that the quantum corrections of the correlator in an off-shell quark state vanish at scales $k^2=-\mu_R^2$ and $k^z=k_R^z$
\begin{eqnarray}\label{RIMOMcondition}
  && Z_{\textrm{OM}}^{-1}(\tau, k_R^z, \mu_R, \Lambda) \left. \langle 0 | (\bar{q} W_{c}) (\tau n_{z}) \slash{n}_{z} \gamma_{5} (W_{c}^{\dagger} h_{v})(0) | b \, \bar q(k) \rangle \right|_{\stackrel{k^2=-\mu_R^2}{k^z=k_R^z}} \nn\\
  &&= \left. \langle 0 |  (\bar{q} W_{c}) (\tau n_{z}) \slash{n}_{z} \gamma_{5} (W_{c}^{\dagger} h_{v})(0) | b \, \bar q(k_R) \rangle \right|_{\textrm{tree}} \,,
\end{eqnarray}
here $\mu_R$ is the renormalization scale. For convenience we would simply denote $\{\tilde \mu\}=\{k^2=-\mu_R^2, k^z=k_R^z\}$ in the rest of the article. It should be stressed that the renormalization condition is applied to the matrix element, not to the quasidistribution itself. In order to get the renormalized quasidistribution, one needs to Fourier transform this matrix element afterwards. The operator in Eq.\,(\ref{RIMOMcondition}) is not $O(4)$ covariant, therefore in addition to $\mu_R$, one needs another scale parameter $k_R^z$ to pin down the renormalization condition. $\Lambda$ denotes the UV cutoff, in the case of dimensional regularization $\Lambda=1/\epsilon$.

We denote the bare correlator for the $B$-meson on the lattice
\begin{eqnarray}
  \tilde{h}_B\left(\tau,k^z,1/\epsilon \right) = \langle 0 | (\bar{q}_{s} W_{c}) \left(\tau n_{z}\right) \slash{n}_{z} \gamma_{5} (W_{c}^{\dagger} h_{v})(0) | b \, \bar q(k) \rangle \,,
\end{eqnarray}
which is renormalized as
\begin{eqnarray}\label{remquasi0}
  \tilde{h}_R\left(\tau,k^z,\{\tilde\mu\} \right) =  Z_{\textrm{OM}}^{-1}(\tau, \{\tilde \mu\}, 1/\epsilon ) \, \tilde{h}_B\left(\tau,k^z,1/\epsilon \right)  \,.
\end{eqnarray}
One advantage of RI/MOM scheme is that although the bare matrix element and the renormalization factor $Z_{\textrm{OM}}$ depend on the choice of regularization scheme, the renormalized matrix element does not. Besides, the logarithmic UV divergence related to self energy of quark and the linear divergence arises from the self energy of Wilson line have been delicately discussed in \cite{Stewart:2017tvs}. All the UV cutoff dependence cancel out in Eq.\,(\ref{remquasi0}) due to the multiplicative renormalizability of quasidistribution amplitude.

Afterwards, by Fourier transforming the renormalized matrix element $\tilde{h}_R\left(\tau,k^z,\{\tilde\mu\} \right)$ to momentum space, one can work out the RI/MOM matching coefficient. This issue will be elaborately discussed in the next section.

\section{Matching between Quasidistribution amplitude and Light-cone Distribution Amplitude}
\label{sec:Calculation}
We now proceed to determine the perturbative matching coefficient that converts the renormalized $B$-meson quasidistribution amplitude in RI/MOM scheme to renormalized  $B$-meson LCDA in $\overline{\textrm{MS}}$ scheme. Following the construction in \cite{Wang:2019msf}, the hard-collinear factorization formula is
\begin{eqnarray}\label{facfor}
  \varphi_{B}^{+}(\xi, \tilde\mu) = \int_{0}^{\infty} d \omega \, H\left( \xi, \omega, n_{z} \!\cdot\! v, \mu, \{\tilde\mu\} \right) \phi_{B}^{+}(\omega, \mu) +\mathcal{O} \left(\frac{\Lambda_{\mathrm{QCD}}}{n_{z} \!\cdot\! v \, \xi}\right) \,.
\end{eqnarray}
For convenience, we subsequently denote $n_{z} \!\cdot\! v$ as $v^z$ in the rest of this paper. The matching coefficient $H$ denotes the difference of UV behavior between the quasiquantity and the light-cone one which is highly nontrivial due to the different presence of the UV cutoff (one can resort to recent reviews \cite{Cichy:2018mum,Ji:2020ect} for more details). But thanks to the asymptotic freedom, this difference can be calculated by perturbation theory in QCD which makes it possible to extract light-cone parton physics from quasiquantities. Notably, the matching coefficient $H$ depends on the choice of renormalization scheme of quasidistribution amplitude.

To determine the matching coefficient at one-loop level, we replace the $B$-meson state with a heavy $b$ quark plus an off-shell light quark state in Eq.\,(\ref{LCDA2}) and Eq.\,(\ref{quasiDA}). Then the matrix elements with the quark state as the initial state can be calculated in perturbation theory. We carry out the calculation using the off-shellness of the light quark as an IR regulator and dimensional regularization with $d=4-2\epsilon$ as the UV regulator.

\begin{figure}[htbp]
	\centering
	\includegraphics[width=1\columnwidth]{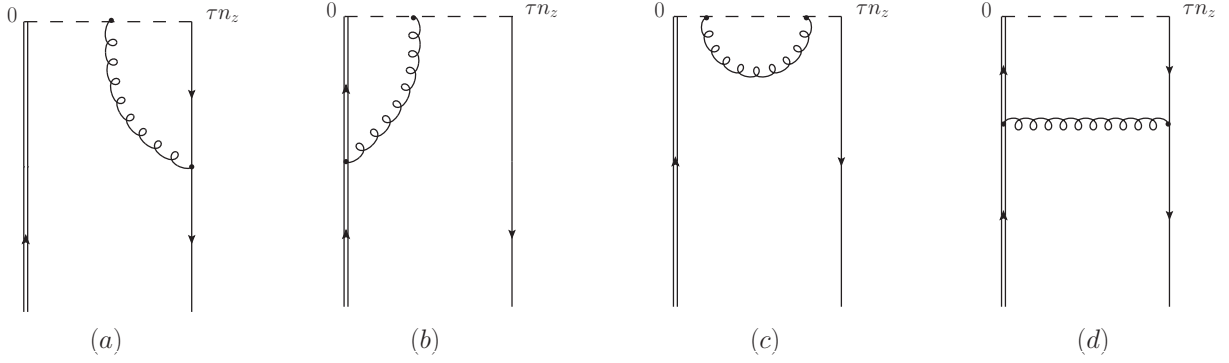}
	\caption{One-loop corrections to the quasidistribution amplitude $\varphi_B^{+}(\xi, \mu) $: the effective HQET  bottom quark is represented by the double line,
 and the spacelike Wilson line is indicated by the dashed line.}
	\label{All-in}
\end{figure}

The one-loop corrections to the quasidistribution amplitude of $\varphi_B^+(\xi,\tilde\mu)$ are shown in Fig.\,\ref{All-in}.
The result at tree level is $\varphi_B^{+(0)}(\xi) = \delta(\xi-\tilde k)$ where $\tilde k\equiv k^z/v^z$. We denote the result of bare quasidsitribution amplitude at one-loop as $\varphi_{B,\textrm{bare}}^{+(1)}(\xi,\mu)$ which has been calculated in \cite{Wang:2019msf}
\begin{align}\label{quasibare}
  \varphi_{B,\textrm{bare}}^{+(1)}(\xi,\mu)
  =&\frac{\alpha_sC_F}{4\pi}\left\{\begin{aligned}
    &\bigg(\frac{1}{\tilde k(\xi-\tilde k)}
    (-\tilde k+2\xi\ln\frac{-\xi}{\tilde k-\xi})\bigg)  +\left[\frac{2}{\tilde k-\xi}\right]_\oplus  \\ &
    +\left[\frac{1}{\xi-\tilde k}\bigg(\frac{1}{\epsilon} -\ln4 +\ln\frac{\mu^2}{v^{z2}(\tilde k-\xi)^2}\bigg)\right]_\oplus
       &\xi<0\\
    & \frac{1}{\tilde k(\xi-\tilde k)}\bigg(2\xi-\tilde k-2\ln\frac{4\tilde k^2v^{z2}}{-k^2}\bigg) + \left[\frac{2}{\tilde k-\xi}\right]_\oplus  \\
    &  +\left[\frac{1}{\xi-\tilde k}\bigg(\frac{1}{\epsilon}  -\ln4 +\ln\frac{\mu^2}{v^{z2}(\tilde k-\xi)^2}\bigg)\right]_\oplus  &0<\xi<\tilde k  \\
    & \frac{1}{\tilde k(\xi-\tilde k)}\bigg(\tilde k-2\xi\ln\frac{\xi}{\xi-\tilde k}\bigg) + \left[\frac{2}{\xi-\tilde k}\right]_\oplus  \\
    &  +\left[\bigg(\frac{1}{\epsilon} +  \ln4 +2\ln v^{z2}+\ln\frac{\mu^2}{v^{z2}(\xi-\tilde k)^2}\bigg)\right]_\oplus  &\xi>\tilde k\\
  \end{aligned}\right.\nn\\
  &+ \frac{\alpha_sC_F}{4\pi} f(a) \, \delta(\xi-\tilde k) \,.
\end{align}
Here, we assign $v^{\mu} = \left(v^0, 0, 0, v^z\right)$ with $v^z\gg 1$. Applying the default power counting scheme one can readily identify that the hard correction from the one-loop box diagram (image (d)) in Fig.\,\ref{All-in} is power suppressed. Recall that we have used the off-shellness of light quark $-k^2$ as IR regulator, this logarithmic IR singularities would cancel between the quasidistribution amplitude and LCDA, leaving the matching coefficient $H$ independent on $-k^2$, as it should be.

The plus distribution is defined by (with $a > 1$)
\begin{eqnarray}\label{plus}
\left \{ {F}(\xi, \omega) \right \}_{\oplus} =  {F}(\xi, \omega)
- \delta(\xi - \omega) \, \int_0^{a \, \xi}  \, d t \,  {F}(\xi, t) \,,
\end{eqnarray}
the subtraction scheme dependent term in Eq.\,(\ref{quasibare})
\begin{eqnarray}
  f(a) &=& -\frac{1}{\epsilon}\Big( 1+\ln\left(4(a-1)v^{z2}\right) \Big) -2-\frac{\pi^2}{3} +4\left( \ln2 \right)^2 +\ln\frac{128}{a-1}+(\ln a-1)^2  \nn\\
  &&+2\ln a +\ln v^{z2}(3+2\ln4+\ln v^{z2}) +\ln(4v^{z2})(3\ln(a-1)-2\ln a) \nn\\
  &&+\operatorname{HPL}[\{-,+\},-1] -2\ln\frac{-k^2}{\tilde k^2}\left(1+\ln\frac{a-1}{a}\right) +\ln\frac{\xi^2}{\mu^2}\Big( 1+\ln(4(a-1)v^{z2}) \Big)
\end{eqnarray}
will compensate the same scheme dependence of the newly introduced  plus distribution for the convolution of the hard function $H$ with a smooth test function. An advantage of introducing the above mentioned plus function is that it allows to implement both the ultraviolet and infrared subtractions for the perturbative matching procedure simultaneously.

Having the bare result at hand, we next discuss the RI/MOM renormalization of $\varphi_B^+(\xi,\tilde\mu)$. The renormalized correlator $\tilde{h}_R\left(\tau,k^z,\{\tilde\mu\} \right)$ has been already given in Eq.\,(\ref{remquasi0}), which is to be Fourier transformed into the $\xi$ space to obtain the distribution $\tilde{\mathcal{F}}(\xi, k^z, \{\tilde\mu\})$:
\begin{eqnarray}
  \tilde{\mathcal{F}}(\xi, k^z, \{\tilde\mu\}) = \int \frac{d \tau}{2 \pi} e^{i v^z \xi \tau} \, \tilde{h}_R\left(\tau,k^z, \{\tilde\mu\} \right) \,.
\end{eqnarray}
$\tilde{\mathcal{V}}( k^z, \{\tilde\mu\})$ is the local correspondence of $\tilde{\mathcal{F}}(\xi, k^z, \{\tilde\mu\})$ which is given by $\tilde{h}_R$ at $\tau=0$,
\begin{eqnarray}
  \tilde{\mathcal{V}}( k^z, \{\tilde\mu\}) = \tilde{h}_R\left(\tau=0,k^z, \{\tilde\mu\} \right) \,.
\end{eqnarray}
With $\tilde{\mathcal{F}}(\xi, k^z, \{\tilde\mu\})$ and $\tilde{\mathcal{V}}( k^z, \{\tilde\mu\})$ calculated on the lattice, the $B$-meson quasidistribution amplitude can be obtained
\begin{eqnarray}\label{phiB_ratio}
  \varphi_B^+(\xi, \tilde\mu) = v^z \int \frac{d \tau}{2 \pi} e^{i v^z \xi \tau} \, \frac{\tilde{h}_R\left(\tau,k^z, \{\tilde\mu\} \right)}{\tilde{h}_R\left(\tau=0,k^z, \{\tilde\mu\} \right)} \,.
\end{eqnarray}

The calculation procedure of the renormalization factor $Z_{\textrm{OM}}$ is similar to the previous one in \cite{Wang:2019msf} but a bit more complicated, since the Feynman diagrams in Fig.\,\ref{All-in} are calculated at a specific scale $\{\tilde\mu\}$. We then proceed to derive the expression of renormalized quasidistribution amplitude $\varphi_B^+(\xi, \tilde\mu)$ from Eq.\,(\ref{phiB_ratio}). Taking advantage of Eq.\,(\ref{remquasi0}), we have
\begin{eqnarray}\label{pre_expand}
  \varphi_B^+(\xi, \tilde\mu) = v^z \int \frac{d \tau}{2 \pi} e^{i v^z \xi \tau} \, \frac{Z_{\textrm{OM}}^{-1}(\tau, \{\tilde \mu\}, 1/\epsilon )}{Z_{\textrm{OM}}^{-1}(0, \{\tilde \mu\}, 1/\epsilon )} ~ \frac{ \tilde{h}_B(\tau,k^z,1/\epsilon) }{ \tilde{h}_B(0,k^z,1/\epsilon) } \,.
\end{eqnarray}
The renormalization constant is determined by the renormalization condition in Eq.\,(\ref{RIMOMcondition})
\begin{eqnarray}\label{ratio_of_ZOM}
  \frac{Z_{\textrm{OM}}^{-1}(\tau, \{\tilde \mu\}, 1/\epsilon )}{Z_{\textrm{OM}}^{-1}(0, \{\tilde \mu\}, 1/\epsilon )} ~ \frac{ \tilde{h}_B(\tau, \{\tilde \mu\} , 1/\epsilon) }{ \tilde{h}_B(0, \{\tilde \mu\} , 1/\epsilon) } &=& \left. \frac{ \tilde{h}_B(\tau, \{\tilde \mu\} , 1/\epsilon) }{ \tilde{h}_B(0, \{\tilde \mu\} , 1/\epsilon) } \right|_{\textrm{tree}} = e^{-i k_R^z \tau} \,,
\end{eqnarray}
in which
\begin{eqnarray}\label{CT_term}
  \frac{ \tilde{h}_B(\tau, \{\tilde \mu\} , 1/\epsilon) }{ \tilde{h}_B(0, \{\tilde \mu\} , 1/\epsilon) } = \int d\xi' e^{-i\tau v^z \xi'} \varphi_{B,\textrm{CT}}^+(\xi', \{\tilde\mu\} ) \,.
\end{eqnarray}
Here $\varphi_{B,\textrm{CT}}^+$ is the additive counterterm contribution of quasidistribution amplitude, as will be clearly seen subsequently. Substitute Eq.\,(\ref{CT_term}) into Eq.\,(\ref{ratio_of_ZOM}), one immediately obtain the ratio of nonlocal and local renormalization constants at one loop
\begin{eqnarray}
  \left( \frac{Z_{\textrm{OM}}^{-1}(\tau, \{\tilde \mu\}, 1/\epsilon )}{Z_{\textrm{OM}}^{-1}(0, \{\tilde \mu\}, 1/\epsilon )} \right)^{(1)} = -\int d\xi' e^{-i\tau(v^z \xi' -k_R^z)} \varphi_{B,\textrm{CT}}^{+(1)}(\xi', \{\tilde\mu\} ) \,,
\end{eqnarray}
as well as $\left(\frac{Z_{\textrm{OM}}^{-1}(\tau, \{\tilde \mu\} )}{Z_{\textrm{OM}}^{-1}(0, \{\tilde \mu\} )}\right)^{(0)} = 1$ at tree level.

Finally, the renormalized quasidistribution amplitude in Eq.\,(\ref{pre_expand}) can be expanded at one-loop order
\begin{eqnarray}\label{counterterm}
  \varphi_B^{+(1)}(\xi, \tilde\mu) &=& v^z \int \frac{d \tau}{2 \pi} e^{i v^z \xi \tau} \,\Bigg\{ \left( \frac{Z_{\textrm{OM}}^{-1}(\tau, \{\tilde \mu\}, 1/\epsilon )}{Z_{\textrm{OM}}^{-1}(0, \{\tilde \mu\}, 1/\epsilon )} \right)^{(1)}   \left(\frac{ \tilde{h}(\tau,k^z) }{ \tilde{h}(0,k^z)}\right)^{(0)} \nn\\
  &&+ \left(\frac{Z_{\textrm{OM}}^{-1}(\tau, \{\tilde \mu\} )}{Z_{\textrm{OM}}^{-1}(0, \{\tilde \mu\} )}\right)^{(0)} \left(\frac{ \tilde{h}_B(\tau,k^z,1/\epsilon) }{ \tilde{h}_B(0,k^z,1/\epsilon)}\right)^{(1)} \Bigg\} \nn\\
  &=& -v^z \int \frac{d \tau}{2 \pi} e^{i v^z \xi \tau} \, \int d\xi' e^{-i\tau(v^z \xi' -k_R^z)} \varphi_{B,\textrm{CT}}^{+(1)}(\xi', \{\tilde\mu\} ) e^{-i k^z \tau}  + \varphi_{B,\textrm{bare}}^{+(1)}(\xi, k^z) \nn\\
  &=& \varphi_{B,\textrm{bare}}^{+(1)}(\xi, k^z) - \varphi_{B,\textrm{CT}}^{+(1)}(\xi+\tilde k_R-\tilde k, r_R ) \,.
\end{eqnarray}
Here $\tilde k_R \equiv k_R^z/v^z$ and we define the dimensionless ratio
\begin{eqnarray}\label{rR}
  r_R \equiv \frac{\mu_R^2}{k_R^{z2}} \,.
\end{eqnarray}
It is worth stressing the difference between $r_R$ and $\rho\equiv -k^2/k^{z2}$. As indicated, we keep $-k^2$ small as the IR regulator, i.e., $\rho\ll 1$. Thus we can identify the logarithmic IR divergences by Taylor expanding in $\rho$, making the calculation much more convenient. However, the renormalization scale $\mu_R$ is not necessarily small, this makes Taylor expansion in $r_R$ unfeasible when calculating the renormalized quasidistribution amplitude, i.e., calculating the counterterm of bare quasidistribution amplitude. More pertinent details on this issue can be found in Appendix.

Next we consider the $B$-meson LCDA $\phi_B^+(\omega,\mu)$ whose IR divergence is regulated by the same light quark off-shellness $-k^2$. With the definition in Eq.\,(\ref{LCDA}), one can get the renormalized $\phi_B^+(\omega,\mu)$ at one-loop in the $\overline{\textrm{MS}}$ scheme:
\begin{align}\label{LCDArem0}
  \phi_{B}^{+(1)}(\omega,\mu)
  =&\frac{\alpha_sC_F}{2\pi}\left\{\begin{aligned}
     & 0 &\omega<0\\
    & \left[
    -\frac{\omega}{(\omega-\tilde k)\tilde k}\ln\frac{\mu^2\tilde k^{2}}{\omega(\tilde k-\omega)(-k^2)}\right]_\oplus  &0<\omega<\tilde k  \\
    & \left[\frac{1}{\omega-\tilde k}\ln\frac{\mu^2}{(\omega-\tilde k)^2}\right]_\oplus  &\omega>\tilde k\\
  \end{aligned}\right. \nn\\
  -& \frac{\alpha_sC_F}{2\pi} \Bigg[ -2+\frac{5\pi^2}{24}+\frac{1}{2}\left(\ln(a-1)\right)^2+(\ln a)^2 +\operatorname{Li}_2(1-a) \nn\\
  +&(\ln a-1)\ln\left(-\frac{\mu^2}{k^2}\right)-\ln(a-1)\ln\left(-\frac{a\mu^2}{k^2}\right)
   +\left(\ln\frac{\omega}{\mu}\right)^2 \Bigg]\delta(\omega-\tilde k) \,.
\end{align}
The results shown in Eq.\,(\ref{LCDArem0}) and Eq.\,(\ref{quasibare}) do not contain the contribution of box diagram, since the collinear contribution to the bare quasidistribution amplitude in box diagram is precisely reproduced by the corresponding diagram for the $B$-meson LCDA at one-loop, i.e., in unphysical region ($\xi<0$), the contribution of box diagram on quasiquantity is suppressed by $1/v_z^2$, and in physical region ($\omega>0$), the contributions on both quasiquantity and LCDA are exactly same. As for the counterterm in box diagram on quasidistribution amplitude in the RI/MOM scheme, as long as we work in the region $v^{z2} \xi \gg 1/r_R$, the contribution can be disregarded. In fact, it has also been demonstrated that the box diagram does not contribute in the pseudo distribution approach neither \cite{Zhao:2020bsx}.

Considering the hard-collinear factorization formula in Eq.\,(\ref{facfor}), the matching coefficient $H$ is then determined by the difference between the momentum space quasiditribution amplitude and LCDA. Expanding $\varphi_{B}^{+}(\xi, \tilde\mu),\, \phi_{B}^{+}(\omega, \mu)$ and $H\left( \xi, \omega, v^z, \mu, \{\tilde\mu\} \right)$ in series of $\alpha_s^n$. Up to one-loop level,
\begin{eqnarray}
	\varphi^+_B(\xi,\tilde\mu)&=&\delta(\xi-\tilde k) + \varphi^{+(1)}_B(\xi,\tilde\mu) + \mathcal{O}(\alpha^2) \,,\nn\\
	 \phi_B^+(\omega,\mu)&=&\delta(\omega-\tilde k) + \phi_B^{+(1)}(\omega,\mu) + \mathcal{O}(\alpha^2) \,,\nn\\
	 H(\xi, \omega, v^z, \mu, \{\tilde\mu\})&=&\delta(\xi-\omega) + H^{(1)}(\xi, \omega, v^z, \mu, \{\tilde\mu\}) + \mathcal{O}(\alpha^2) \,.
\end{eqnarray}
Substituting the expressions above into Eq.\,(\ref{facfor}),
\begin{eqnarray}\label{calofH1}
	\left. H^{(1)}(\xi, \omega, v^z, \mu, \{\tilde\mu\})\right|_{\omega\to \tilde k} = \varphi^{+(1)}_B(\xi,\tilde\mu)- \left. \phi_B^{+(1)}(\omega,\mu)\right|_{\omega\to\xi} \,.
\end{eqnarray}
The renormalized $\varphi^{+(1)}_B(\xi,\tilde\mu)$ and $\phi_B^{+(1)}(\omega,\mu)$ have already been calculated, therefore the matching coefficient can be derived from Eq.\,(\ref{calofH1}),
\begin{eqnarray}\label{CoeH}
   H(\xi, \omega, v^z, \mu, \{\tilde\mu\}) = \delta(\xi-\omega) + g_1(\xi,\omega,\mu) -g_2(\xi,\omega,\{\tilde \mu\}) +\frac{\alpha_sC_F}{4\pi}\ln v^z \left(3+4\ln\frac{a-1}{a}\right)\delta(\xi-\omega)  \,, \nn\\
\end{eqnarray}
where
\begin{align}\label{g1}
  g_1(\xi, \omega, \mu)
  =&\frac{\alpha_sC_F}{4\pi}\left\{\begin{aligned}
     & \frac{1}{\omega(\omega-\xi)}\left( \omega-2\xi\ln\frac{-\xi}{\omega-\xi} \right) &\xi<0\\
    & \left[\frac{1}{\omega( \omega-\xi )}\left( \omega -2\xi +2\xi \ln\frac{4v^{z2}\xi(\omega-\xi)}{\mu^2}  \right) \right]_\oplus  &0<\xi<\omega \\
    & \left[\frac{1}{\omega (\omega-\xi )}\left(-\omega +2 \omega \ln \frac{\mu^2}{(\xi-\omega)^2} +2\xi\ln\frac{\xi}{\xi-\omega}\right) \right]_\oplus  &\xi>\omega\\
  \end{aligned}\right.  \,,
\end{align}
and
\begin{align}\label{g2}
  g_2(\xi, \omega, \{\tilde\mu\})
  =&\frac{\alpha_sC_F}{4\pi}\left\{\begin{aligned}
     & -\frac{1}{\omega-\xi} &\xi<\omega-\tilde k_R \\
    & \Bigg[ \frac{1}{2k_R^z\sqrt{1-r_R}(\omega-\xi)} \bigg( -2\sqrt{1-r_R}(k_R^z+2v^z(\xi-\omega)) \\ &
    -\big(4v^z(\xi-\omega)-k_R^z(r_R-4)\big)\ln\frac{2-2\sqrt{1-r_R}-r_R}{r_R}  \bigg) \Bigg]_\oplus  & \omega-\tilde k_R<\xi<\omega \\
    & \left[ \frac{1}{\omega-\xi} \right]_\oplus  &\xi>\omega\\
  \end{aligned}\right.  \,,
\end{align}
As expected, $H$ does not depend on the IR regulator $-k^2$ since the logarithmic IR singularities cancel between the quasidstribution amplitude and the LCDA. The $\mathcal{O}(1/v^{z2})$ contributions to the matching coefficient $H$ are dropped, the $v^z$ expansion is subtle thus should be treated carefully and systematically. One can tell that the expression of $H$ is more complicated than the one in \cite{Wang:2019msf} where the quasidistribution amplitude is renormalized in the $\overline{\textrm{MS}}$ scheme, this is natural since the renormalization condition in RI/MOM has introduced new momentum scales $\{\tilde \mu\}$. In Sec.\,\ref{sec:Numerical} we will make comparison between these two matching coefficients.

\section{Perspectives for Lattice Calculations}
\label{sec:Numerical}
\begin{figure}[htbp]
\centering
\includegraphics[width=0.60\columnwidth]{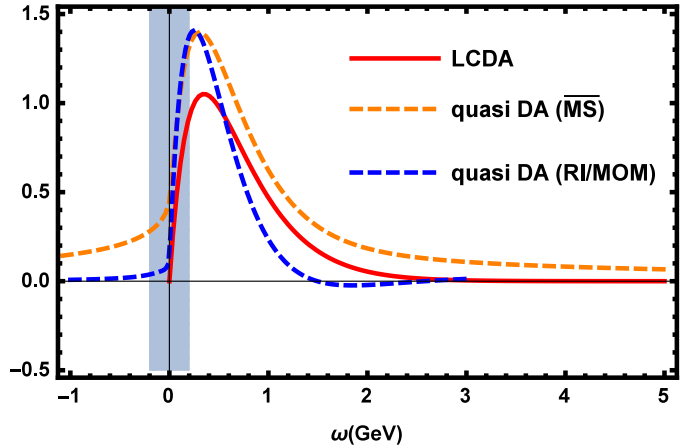}
\centering
\caption{The shapes of the $B$-meson quasidistribution amplitude $\varphi_B^{+}(\xi\!=\!\omega, k_R^z\!=\!2.0 \, {\rm GeV}, r_R \!=\! 2)$ in bHQET obtained from the hard-collinear factorization theorem in Eq.\,(\ref{facfor}) and from the nonperturbative model of $\phi_{B}^{+}(\omega, \mu\!=\!1.5 \, {\rm GeV})$ presented in Eq.\,(\ref{modelLCDA}). The red solid line represents the nonperturbative model of $\phi_{B}^{+}$, the corresponding quasidistribution amplitudes $\varphi_{B}^{+}$ normalized in the $\overline{\textrm{MS}}$ and RI/MOOM schemes are presented respectively (orange dashed and blue dashed lines). The shadow region of $|\omega| \leq 200 ~{\rm MeV}$ is excluded due to inapplicability of the hard-collinear factorization formula for $|v^z \omega| \leq 2.0 ~{\rm GeV}$.}
\label{diagram2}
\end{figure}

We discuss the perspectives for lattice calculations based on numerical analysis. An important step in obtaining the $B$-meson LCDA in bHQET based upon LaMET is to perform the lattice simulation of the quasidistribution amplitude $\varphi_B^+(\xi,\tilde\mu)$ in the moving $B$-meson frame with $v^z\gg 1$. To this end, it will be instructive to study how the matching coefficient in Eq.\,(\ref{CoeH}) changes the LCDA, helping people understand the characteristic feature of $\varphi_B^+(\xi,\tilde\mu)$. We start with a well known phenomenological model of $\phi_B^+(\omega,\mu)$ motivated by the HQET sum rule calculation \cite{Grozin:1996pq}
\begin{eqnarray}\label{modelLCDA}
  \phi_{B}^{+}(\omega, \mu=1.5~\mathrm{GeV})=\frac{\omega}{\omega_{0}^{2}} e^{-\omega / \omega_{0}} \,,
\end{eqnarray}
here the reference value of the logarithmic inverse moment $\omega_0=350 ~\mathrm{MeV}$ is taken for illustration purposes. With the expression of $\phi_B^+(\omega,\mu)$ above and the factorization formula in Eq.\,(\ref{facfor}), we can depict the shape of quasidistribution amplitude $\varphi_B^+(\xi,\tilde\mu)$. For our study, we set the default values $k_R^z=2 ~\textrm{GeV},\, \mu=1.5 ~\textrm{GeV},\,  r_R=2$. The factorization formula in Eq.\,(\ref{facfor}) requires a large $v^z$ in order to suppress the $\mathcal{O}(1/v^{z2})$ corrections, here we take $v^z=10$. Fig.\,\ref{diagram2} shows comparisons between the RI/MOM quasidistribution amplitude (blue dashed line), the $\overline{\textrm{MS}}$ quasidistribution amplitude (orange dashed line) and the modeled LCDA of $B$-meson (red solid line). One can see that both RI/MOM and $\overline{\textrm{MS}}$ quasidistribution amplitudes are close to the $B$-meson LCDA, and the radiative tail at large and negative momentum $\xi$ developed in $\overline{\textrm{MS}}$ quasidistribution amplitude does not emerge in RI/MOM quasidistribution amplitude, which is encouraging on account of the convergence of perturbation theory in RI/MOM scheme. In addition, in contrast to the quasiparton distribution function in \cite{Stewart:2017tvs}, no peaks arise in the momentum region $\xi\leq0$.

\begin{figure}[htbp]
\centering
\includegraphics[width=0.49\columnwidth]{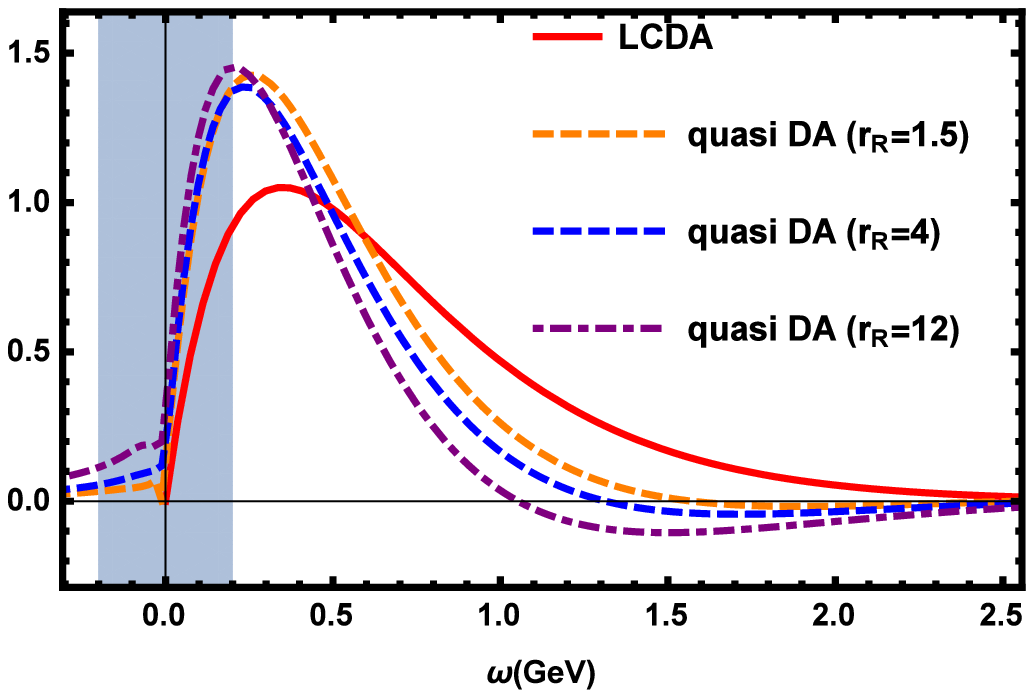}
\includegraphics[width=0.49\columnwidth]{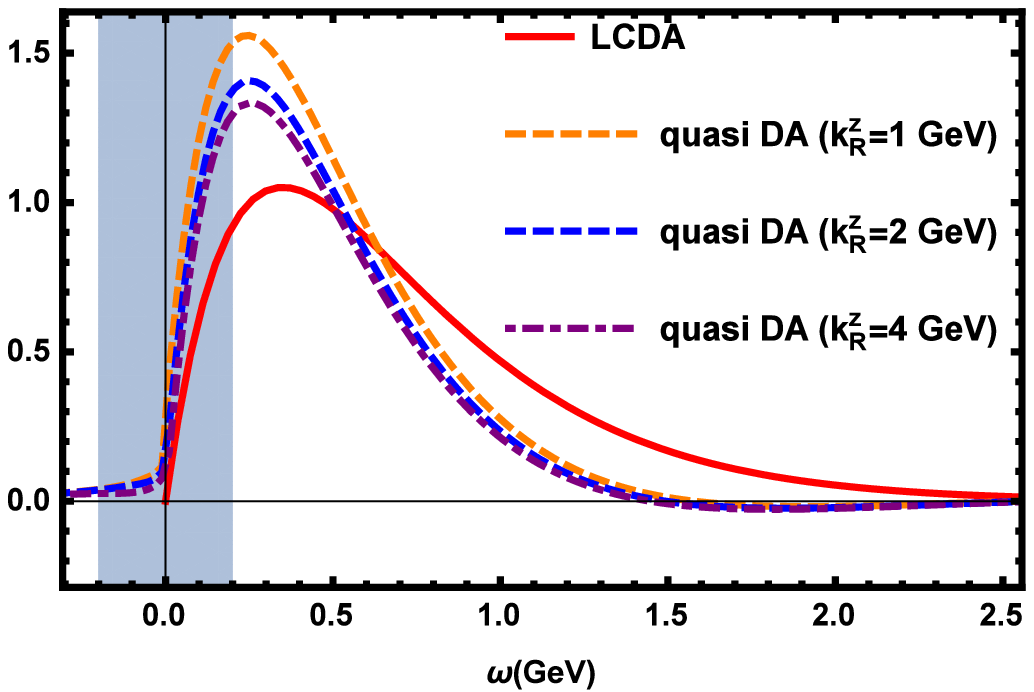}
\centering
\caption{Comparisons between the LCDA and the quasidistribution amplitude obtained at different $r_R$s (left panel) and $k_R^z$s (right panel).}
\label{diagram23}
\end{figure}

Next we consider the dependence of RI/MOM quasidistribution amplitude on $r_R$ and $k_R^z$. We fix $k_R^z=2 ~\textrm{GeV}, \mu=1.5 ~\textrm{GeV}, v^z=10$ and vary the parameter $r_R=\{1.5,\, 4,\, 12\}$ in the left panel of Fig.\,\ref{diagram23}. One can tell the quasidistribution amplitude is pretty sensitive to the variation of $r_R$. It seems that with larger $r_R$, the quasidistribution amplitude moves away from LCDA. In the right panel of Fig.\,\ref{diagram23} we vary $k_R^z=\{1,\, 2,\, 4\}~\textrm{GeV}$ with fixed value of $r_R=2, \mu=1.5 ~\textrm{GeV}, v^z=10$.

\begin{figure}[htbp]
\centering
\includegraphics[width=0.60\columnwidth]{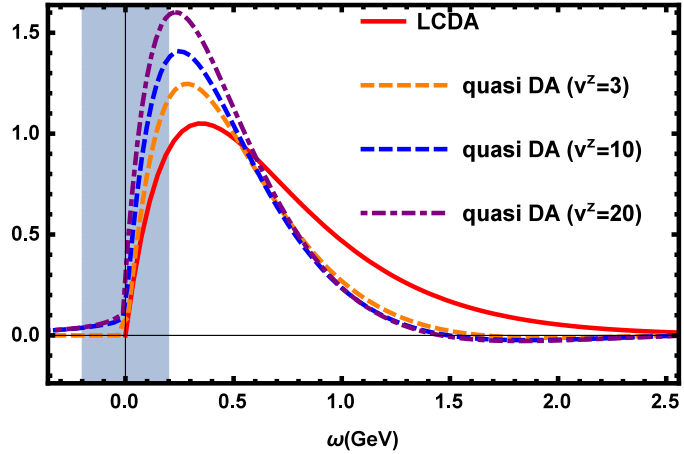}
\centering
\caption{Comparisons between the LCDA and the quasidistribution amplitude obtained at $v^z=3$ (orange dashed), $v^z=10$ (blue dashed) and $v^z=20$ (purple dashed).}
\label{diagram4}
\end{figure}

Finally we discuss the dependence on the heavy quark velocity $v^z$. We hold $k_R^z=2 ~\textrm{GeV}, \mu=1.5 ~\textrm{GeV}, r_R=2$ and vary $v^z=\{3,\, 10,\, 20\}$ in Fig.\,\ref{diagram4}. The differences between quasidistribution amplitudes depicted with different $v^z$ reduce rapidly as $\omega$ increases. When $\omega>0.8$, the three lines almost merged into one, simliar feature has also been observed in the study of quasiparton distribution function depicted with different $P^z$ \cite{Stewart:2017tvs}, which suggests the RI/MOM scheme a promising approach with favourable convergence at large $\omega$.

In conclusion, the numerical analysis in this section indicates that the RI/MOM scheme is suitable for renormalizing $B$-meson quasidistribution amplitude. The derived one-loop matching coefficient yields only a relatively small effect on the modeled $B$-meson LCDA, which bring more confidence about extracting $B$-meson LCDA perturbatively and model-independently in the future. It should be stressed here that our major objective is to explore the opportunity of accessing the light-cone dynamics of the $B$-meson leading-twist distribution amplitude by simulating the RI/MOM quasidistribution amplitude on the lattice, it is a rather preliminary attempt. Actually, the numerical simulations of such quasidistribution amplitudes are still at an exploratory stage, even for the ones suitable for the determination of the light-meson distribution amplitude. Improved methodologies to control both the statistical errors and the systematic uncertainties are called for, as well as further development of new algorithms and computing techniques on the lattice (see \cite{Ji:2020ect,Cichy:2018mum,Alexandrou:2019lfo,Zhang:2019qiq} for details on lattice calculation). We would also like to remind the readers here that a hybrid renormalization procedure has been proposed for quasiparton distribution function recently, which utilizes the advantages of RI/MOM and ratio schemes at short and large distances simultaneously \cite{Ji:2020brr}. The study of the feasibility of this renormalization procedure applied to $B$-meson quasidistribution amplitude deserves more attention.

\section{Conclusion}
\label{sec:Conclusion}
LaMET theory has provided a natural way to calculate parton distributions in an interval of momentum scales, similar to extracting parton distributions from experimental data at finite energies. Within the framework of LaMET, we have derived the matching coefficient which connects the renormalized quasiditribution amplitude in the RI/MOM scheme and standard LCDA in the $\overline{\textrm{MS}}$ scheme. Our numerical analysis indicates that the one-loop matching has nice UV convergence and reasonable magnitude as a perturbative correction, which shows the theoretical uncertainty caused by perturbative matching is controllable, thus making the RI/MOM scheme feasible in lattice applications. We believe that our result has the potential to considerably improve the convenience and accuracy of extracting $B$-meson LCDA from quasiquantities, hence to promote the development for the first-principle determination of the highly desired $B$-meson LCDA, which is undoubtedly of the highest importance for exploring the delicate flavor structure of the SM and beyond at LHCb and Belle II experiments.

To further increase the accuracy of our results, one can study the yet unavailable higher order perturbative corrections to the short-distance matching coefficient and construct the subleading-power factorization formula for the quasidistribution amplitude, which we would like to leave for future works.

\section*{Acknowledgements}
We thank Prof. Wei Wang for careful reading of the manuscript and suggestions. We are also grateful to Dr. Shuai Zhao for inspiring discussions and valuable comments on the renormalization process of $B$-meson quasidistribution amplitude. J.X. and X.R.Z are supported by National Natural Science Foundation of China under Grant No. 12105247 and 12047545, the China Postdoctoral Science Foundation under Grant No. 2021M702957.

\appendix
\section{Renormalization of $B$-meson quasidistribution amplitude}
First, consider the amplitude of heavy-quark sail diagram (image (b) in Fig.\,\ref{All-in})
\begin{eqnarray}\label{pair1}
	\varphi_{B,\textrm{bare}}^{+(b)}(\xi, \mu) &=& i g_{s}^{2} C_{F} \tilde{\mu}^{2 \epsilon} v^z \int \frac{d^{d} q}{(2 \pi)^{d}} \frac{1}{q^{z}} \frac{1}{q^{2}} \frac{1}{v \cdot q}  \Big( \delta(\xi-\tilde k+q^z)-\delta(\xi-\tilde k ) \Big) \,.
\end{eqnarray}
Here the delta functions $\delta(\xi-\tilde k+q^z)$ and $\delta(\xi-\tilde k )$ in the parenthesis in Eq.\,(\ref{pair1}) come from the Fourier transformation with respect to the variable $\tau$ in the ``real'' and ``virtual'' diagrams respectively. Notably, all the $k$ dependence comes from delta function, the other part of the integrand is independent on $k^z$ or $\rho\equiv -k^2/k^{z2}$, indicating the corresponding counterterm $\varphi_{B,\textrm{CT}}^{+(b)}(\xi+\tilde k_R-\tilde k, r_R )$ in Eq.\,(\ref{counterterm}) remains unchanged when the RI/MOM renormalization condition is imposed at the specific scale $\{\tilde\mu\}$,
\begin{eqnarray}\label{1pair1}
	\varphi_{B,\textrm{CT}}^{+(b)}(\xi+\tilde k_R-\tilde k, r_R ) &=& i g_{s}^{2} C_{F} \tilde{\mu}^{2 \epsilon} v^z \int \frac{d^{d} q}{(2 \pi)^{d}} \frac{1}{q^{z}} \frac{1}{q^{2}} \frac{1}{v \cdot q}  \Big( \delta(\xi-\tilde k+q^z)-\delta(\xi-\tilde k ) \Big) \,.
\end{eqnarray}
Therefore, the contribution of heavy-quark sail diagram cancels out after renormalization. This feature which raises in the RI/MOM $B$-meson quasiditribution amplitude considerably simplifies our calculation and facilitates a relatively small effect of the final one-loop matching coefficient. Similar cancelation also appears in the Wilson line self-energy diagram (image (c) in Fig.\,\ref{All-in}),
\begin{eqnarray}\label{1pair2}
	\varphi_{B,\textrm{bare}}^{+(c)}(\xi, \mu) &=& -i g_{s}^{2} C_{F} \tilde{\mu}^{2 \epsilon} \int \frac{d^{d} q}{(2 \pi)^{d}}  \frac{1}{q^2}\frac{1}{q^{z2}}  \Big( \delta(\xi-\tilde k+q^z)-\delta(\xi-\tilde k) \Big) \,.
\end{eqnarray}
Once again, the integrand except for the delta function in Eq.\,(\ref{1pair2}) is independent on $k^z$ or $\rho$, indicating the contributions of the bare term and counterterm cancel out after RI/MOM renormalization.

As for the box diagram (image (d) in Fig.\,\ref{All-in}), the result of bare quasidistribution amplitude reads
\begin{eqnarray}
   \varphi_{B,\textrm{bare}}^{+(d)}(\xi, \mu) &=& \frac{\alpha_{s} C_{F}}{2\pi}\left\{ -\frac{\xi}{\tilde k(\tilde k-\xi)}\ln\frac{\xi}{\xi-\tilde k}\theta(\xi-\tilde k) +\frac{\xi}{\tilde k(\tilde k-\xi)}\ln\frac{-k^2}{\tilde k^2}\theta(0<\xi<\tilde k)+ (0) \, \theta(\xi<0)  \right\} \nn\\
  && +\mathcal{O}\left(1/v^{z2}\right) \,.
\end{eqnarray}
It is worth noting that the contribution to the bare quasidistribution amplitude in box diagram at physical region ($\theta(\xi-\tilde k)$ and $\theta(0<\xi<\tilde k)$) is exactly same as the corresponding box diagram for the $B$-meson LCDA, and the contribution at unphysical region ($\xi<0$) is suppressed by $1/v_z^2$ (the contribution of $B$-meson LCDA at unphysical region is $0$). Besides, the box diagram does not introduce any UV divergence, therefore despite of its intricate form, the corresponding counterterm on quasidistribution amplitude provides only finite terms which is of order $O(1/v^{z2})$. Summarize the above, the box diagram does not contribute to the matching coefficient within $\mathcal{O}(1/v^{z0})$ accuracy. In fact, it has already been shown in \cite{Wang:2019msf,Xu:2022krn} that the box diagram does not contribute both in the LaMET and pseudo distribution approaches.

Finally we consider the light-quark sail diagram (image (a) in Fig.\,\ref{All-in}). Write down the expression for the bare quasidistribution amplitude
\begin{eqnarray}\label{lightsail}
  \varphi_{B,\textrm{bare}}^{+(a)}(\xi, \mu) &=& -i g_{s}^{2}C_F(\tilde{\mu})^{2 \epsilon} \int \frac{d^{d} q}{(2 \pi)^{d}} \frac{1}{q^{z}} \frac{1}{q^{2}} \frac{1}{(q+k)^{2}} \left(k^z(\rho-2)-q^z-q^t\sqrt{1-\rho}\right)  \nn\\
  &&\times \Big( \delta(\xi-\tilde k-q^{z}) - \delta(\xi-\tilde k)  \Big) \,.
\end{eqnarray}
We have utilized a projection operator to deal with the Dirac matrix $\bar{v}(k) \Gamma u_{v}\left(p_{b}\right) \rightarrow \operatorname{Tr}\left[\frac{1+\slashed{v}}{2} M_b\slashed{v} \gamma_{5} \slashed{k} \Gamma\right]$. In addition to the delta function, the other part of the integrand in Eq.\,(\ref{lightsail}) has $k$ dependence. The result of bare amplitude reads has already been calculated in \cite{Wang:2019msf}.

As for the counterterm in RI/MOM scheme, it is determined by setting $k^2=-\mu_R^2$ and $k^z=k_R^z$.
\begin{eqnarray}\label{lightsailCT}
  \varphi_{B,\textrm{CT}}^{+(a)}(\xi+\tilde k_R-\tilde k, r_R ) &=& -i g_{s}^{2}C_F(\tilde{\mu})^{2 \epsilon} \int \frac{d^{d} q}{(2 \pi)^{d}} \frac{1}{q^{z}} \frac{1}{q^{2}} \frac{1}{(q+k_R)^{2}} \left(k_R^z(r_R-2)-q^z-q^t\sqrt{1-r_R}\right)  \nn\\
  &&\times \Big( \delta(\xi-\tilde k-q^{z}) - \delta(\xi-\tilde k)  \Big) \,.
\end{eqnarray}
The $r_R$ is not necessarily small, this makes Taylor expansion in $r_R$ unfeasible in calculation. After introducing Feynman parameter $\alpha$ and integrating the $d-1$ dimensions of integral momentum $q$, we have
\begin{eqnarray}
  \varphi_{B,\textrm{CT}}^{+(a)}(\xi+\tilde k_R-\tilde k, r_R ) &=& -\frac{2\alpha_sC_F}{8\pi^{\frac{3}{2}}} \int_{0}^{1}\!\!d\alpha\!\! \int_{-\infty}^{+\infty}\!\!dq^z  \frac{e^{\gamma_E\epsilon}\mu^{2\epsilon}(q^z+k_R^z(2+\alpha(r_R-1)-r_R))\Gamma(\frac{1}{2}+\epsilon)}{q^z \left( q^{z2}+2k_R^z q^z \alpha+k_R^{z2}\alpha(\alpha+r_R-\alpha r_R) \right)^{\frac{1}{2}+\epsilon}}\nn\\
  &&\times  \Big( \delta(\xi-\tilde k-q^z)-\delta(\xi-\tilde k) \Big) \,.
\end{eqnarray}
Subsequently we integrate out $\alpha$ and $q^z$ and get the result of this counterterm which will be incorporated into the final result of renormalized quasidistribution amplitude in Eq.\,(\ref{remquasi}) below.

With all these results shown above at one-loop, the renormalized quasidistribution amplitude can be written down,
\begin{eqnarray}\label{remquasi}
  \varphi_{B}^{+}(\xi, \tilde\mu) &=& \delta(\xi-\tilde k) + h_1(\xi, \tilde k) - h_2(\xi, \{\tilde\mu\}) +\frac{\alpha_sC_F}{4\pi}\ln v^z \left(3+4\ln\frac{a-1}{a}\right)\delta(\xi-\tilde k) \,,
\end{eqnarray}
where
\begin{align}\label{h1}
  h_1(\xi, \tilde k)
  =&\frac{\alpha_sC_F}{4\pi}\left\{\begin{aligned}
     & \frac{1}{\tilde k(\xi-\tilde k)}\left( -\tilde k+2\xi\ln\frac{-\xi}{\tilde k-\xi} \right) &\xi<0\\
    & \left[\frac{1}{\tilde k( \xi-\tilde k )}\left(2 \xi-\tilde k-2 \xi \ln \frac{4\tilde k^{2} v^{z2}}{-k^2}\right) \right]_\oplus  &0<\xi<\tilde k  \\
    & \left[\frac{1}{\tilde k (\xi-\tilde k )}\left(\tilde k-2 \xi \ln \frac{\xi}{\xi-\tilde k}\right) \right]_\oplus  &\xi>\tilde k\\
  \end{aligned}\right.  \,,
\end{align}
and
\begin{align}\label{h2}
 h_2(\xi, \{\tilde\mu\})
  =&\frac{\alpha_sC_F}{4\pi}\left\{\begin{aligned}
     & \frac{1}{\xi-\tilde k} &\xi<\tilde k-\tilde k_R \\
    & \Bigg[ \frac{1}{2k_R^z\sqrt{1-r_R}(\tilde k-\xi)} \bigg( -2\sqrt{1-r_R}(k_R^z+2v^z(\xi-\tilde k))   \\ &
    + \big(k_R^z(r_R-4)+4v^z(\tilde k-\xi)\big)\ln\frac{2-2\sqrt{1-r_R}-r_R}{r_R} \bigg) \Bigg]_\oplus  &\tilde k-\tilde k_R<\xi<\tilde k  \\
    & \left[ -\frac{1}{\xi-\tilde k} \right]_\oplus  &\xi>\tilde k\\
  \end{aligned}\right.  \,.
\end{align}

Bringing the renormalized quasidistribution amplitude $\varphi_{B}^{+}(\xi, \tilde\mu)$ in Eq.\,(\ref{remquasi}) and the renormalized LCDA $\phi_B^{+}(\omega,\mu)$ in Eq.\,(\ref{LCDArem0}) into Eq.\,(\ref{calofH1}), we get the expected matching coefficient in Eq.\,(\ref{CoeH}).

\end{document}